# Universal scaling of latent heat of orbital order-disorder transition with average R-site ion size in perovskite $RMnO_3$ systems

Parthasarathi Mondal and Dipten Bhattacharya*
*Electroceramics Division, Central Glass and Ceramic Research Institute, Calcutta 700 032, India*

The latent heat ($L$) of orbital order-disorder transition in single-valent perovskite manganite series $La_{1-x}R_xMnO_3$ (R = Pr, Nd, Gd; x = 0.0-1.0) decreases with the decrease in average R-site radius $<r_R>$ following a *universal scaling law* $L.<r_R>^2/\sigma^2 \sim \exp(<r_R>)$, where $\sigma^2$ is the variance in R-site radius, and eventually reaches zero at a *critical R-site radius* $<r_R>_c \approx 1.180$ Å. Such a drop in $L$ is due, possibly, to a universal pattern of evolution of finer orbital domain structure with the drop in $<r_R>$ as well as with the increase in $\sigma^2$ irrespective of R-site ion type.

Different orbital phases have been identified both theoretically and experimentally in a series of strongly correlated electron systems.[1] They range from orbital solid, liquid, glass, liquid crystal to more esoteric phases in two-dimensional triangular lattice ($LiVO_2$) or one dimensional TiOCl, charge transfer insulators, or spinels with transition metal ions having partially filled $t_{2g}$ levels at B-sites as in $NaTiO_2$ (orbitally induced Peierls phase) etc.[2-6] The study of evolution of one phase to another is quite fascinating and rewarding as it provides insights into the fine interplay among orbital, spin, charge, and lattice degrees of freedom. It also offers clues for newer micro-electronic applications based on orbitronics.[7] One of the techniques commonly employed in characterizing an orbital phase and its evolution is the study of its phase transition characteristics. Recent results[8-12] in this regard provide very interesting picture. The first order transition, characterized by finite latent heat and lattice volume contraction, gives way to second order transition as lattice distortion increases. More direct characterization of the orbital phases using local structural data provides information about the orbital domain size and its temporal fluctuations. As lattice distortion increases, the orbital domains undergo spatio-temporal evolution with decrease in size and increase in temporal fluctuation. This, in turn, leads to evolution in the order of the transition. For few selected undoped and doped perovskite manganite compounds like $LaMnO_3$, $La_{0.8}Ca_{0.2}MnO_3$, $Pr_{0.6}Ca_{0.4}MnO_3$, the orbital domain size has been estimated.[13,14] However, there is, as yet, no systematic mapping of spatio-temporal fluctuation of the orbital domains available across the entire phase diagram of strongly correlated electron system as a function of evolution of phase transition order. Neither is there any systematic mapping of the evolution of the phase transition as a function of lattice distortion. In this paper, we attempt to provide, at least, a mapping of variation of latent heat ($L$) associated with the orbital order-disorder transition as a function of average R-site ion radius ($<r_R>$) in a series of undoped perovskite $RMnO_3$.

We report that $L$ decreases with the decrease in $<r_R>$ and reaches zero at a critical $<r_R>_c \approx 1.180$ Å. We also report that $L$ depends on variance $\sigma^2$ as well, especially, around the zone where $\sigma^2 \rightarrow \sigma^2_{max}$. It is possible to show that the modified $L.<r_R>^2/\sigma^2$ exhibits a universal scaling with $<r_R>$.

The experiments have been carried out on high quality bulk polycrystalline samples. They are prepared by compacting the powder, obtained through solution chemistry route[11] and subsequent calcination and sintering under inert (Ar) atmosphere at 1300-1400$^o$C for 5-10 h. The samples have been characterized by X-ray diffraction (XRD), energy dispersive X-ray (EDX), scanning electron microscopy (SEM), chemical analysis for estimation of $Mn^{4+}$ concentration, dc resistivity measurements etc. The calorimetric measurements have been carried out in differential thermal analyzer (DTA) of Shimadzu, Japan and in differential scanning calorimeter (DSC) of Perkin-Elmer (Pyris Diamond DSC). The dc resistivity has been measured over a temperature regime 300-1200 K using high quality platinum paste and wires. The contacts were cured overnight at 1400 K under inert atmosphere.

The $Mn^{4+}$ concentration is found to vary within 2-4% in all the cases. The estimation of room temperature lattice parameters, orthorhombic distortion ($D$), and volume ($V$) reveals systematic patterns of variation with the decrease in $<r_R>$ (Fig. 1). It is interesting to note that, at room temperature, both $V$ and $D$ of different systems follow a rather universal pattern of variation with $<r_R>$ with a small scattering due to difference in R-site ion type – La, Pr, Nd, Gd. For comparison, the patterns of variation of $V$ and $D$ with the doping level ($x$) are shown in Fig. 1. The $V$, $D$ versus $x$ pattern depends clearly on the R-site ion type.

In Fig. 2, we show the resistivity vs. temperature plots, and the DTA/DSC thermograms. The transition temperatures ($T_{JT}$), obtained from these studies, corroborate each other and confirm the consistency in the quality of the samples. The latent heat ($L$) is calculated from the DSC thermograms by properly subtracting the background. In Fig. 2c, $L$ is shown as a function of $x$ for all the ions – Pr, Nd, and Gd – while the $T_{JT}$ versus $x$ patterns are shown in Fig. 2d. $L$ is found to drop with $x$ following different patterns depending on the type of ion: (i) for Pr, it drops sharply within $x$ = 0.0-0.2, then a rather flat pattern over $x$ = 0.2-0.7, and finally a sharp drop for $x$ = 0.7-1.0; (ii) for Nd, it drops sharply within $x$ = 0.0-0.3 and then gradually over $x$ = 0.3-0.7; (iii) for Gd, it drops sharply to zero within $x$ = 0.0-0.33. $T_{JT}$, as expected, rises linearly with $x$ for each series with ion dependent slope $dT_{JT}/dx$: ~33 for Pr, ~40 for Nd, and ~39 for Gd. The reason behind different patterns of drop in $L$ with $x$ for different ions could be the doping induced disorder $\sigma^2$, which varies from ion to ion. The slight decrease in $dT_{JT}/dx$ for Gd-doped samples could be due to marginally higher $Mn^{4+}$ content in those samples. In Fig. 3, we show that suitable choice of the scaling parameters helps in collapsing the set of $L$ versus $x$ patterns. The plot of $L.<r_R>^2/\sigma^2$ versus $<r_R>$ depicts a universal pattern.

Therefore, it appears that irrespective of the R-site ion type the crossover from first to higher order transition of the orbital phase depends only on $\langle r_R \rangle$ and $\sigma^2$. *This is the central result of this work.*

It is clear from the entire set of crystallographic and calorimetric measurements that the doping induced disorder due to mismatch in R-site ion size does influence the latent heat (*L*) although it fails to influence the global structural parameters such as lattice volume (*V*) or orthorhombic distortion (*D*). The role of $\sigma^2$ is subtle. We can rationalize these observations by resorting to a concept of variation in orbital domain size depending on overall lattice distortion due both to $\langle r_R \rangle$ and $\sigma^2$. The orbital domain sizes have been estimated[13,14] for $LaMnO_3$, $La_{0.8}Ca_{0.2}MnO_3$, and $Pr_{0.6}Ca_{0.4}MnO_3$ systems and are found to be varying between 4000-400 Å with the decrease in domain size in the case of doped systems as well as for systems with smaller R-site radius. These data give an idea that the orbital domain size does decrease as doping level increases and/or $\langle r_R \rangle$ decreases. The evolution of L to zero at $\langle r_R \rangle_c$ shows that irrespective of ion type at R-site, orbital domain size in such cases reaches a certain universal value that offers *L* comparable to or below the instrument resolution in *global* calorimetry. Only a high precision *local* calorimetry might resolve the latent heat of transition in such cases.[15] The increase in disorder too leads to decrease in orbital domain size due, possibly, to segregation of structurally dissimilar phases. It turns out that *L* is more sensitive to the variation in orbital domain size than the global crystallographic parameters *V* and *D*.

The interplay of Jahn-Teller orbital order with $GdFeO_3$-type tilt order in $RMnO_3$ has earlier been shown[16] to give rise to d-type orbital order for those cases where tilt is large due to smaller $\langle r_R \rangle$. This results from covalency in R-O bonds. In a recent model[17] it has been shown how elastic near-neighbor and next-near-neighbor interaction among the Jahn-Teller ions and/or impurities stabilizes different superstructures including orbital domains. The strain field ($E$) due to impurities or Jahn-Teller ions has long-range ($r$) effect as $E \sim 1/r^3$ [Ref. 18]. As a result, coupling of defects and/or Jahn-Teller ions develop for formation of different orbital order superstructures. In this scenario, it is possible to envisage a model of interference of strain fields originating from different sources: (i) Jahn-Teller ions, (ii) R-site mismatch which, in turn, gives rise to tilt in $Mn^{3+}O_6$ octahedra, and (iii) phase segregation due to large $\sigma^2$. Such interference will limit the growth of orbital order and will give rise to orbital domains. Finally, the domain boundary energy will determine the orbital domain size. Since $L$ depends directly on the orbital domain size, universal pattern of drop in $L$ with $\langle r_R \rangle$ and $\sigma^2$ signifies that the orbital domain size also decreases following a universal pattern. The interesting point here is that since such orbital domain structure does not depend on R-site ion type, it is possible to design an orbital domain structure by R-site ion type invariant choice of $\langle r_R \rangle$ and $\sigma^2$. This will certainly bring more flexibility in designing a domain structure for orbitronics-based applications. The quantitative estimation of universal evolution pattern of orbital domain structure as a function of $\langle r_R \rangle$ and $\sigma^2$ will be attempted in future.

In summary, we report a rather universal pattern of variation in latent heat of orbital order-disorder transition in perovskite $RMnO_3$ series with the average R-site ion size. It has been found that irrespective of R-site ion type the latent heat becomes zero at a critical R-site size $<r_R>_c$ ~ 1.180 Å. The role of doping induced disorder on the latent heat is palpable, especially, where it attains maximum. Such a pattern could possibly be due to a universal pattern of evolution of orbital domain structure as a function of both $<r_R>$ and $\sigma^2$. Our observation creates a possibility of tuning the orbital order melting with desired latent heat at a desired temperature through subtle engineering with $<r_R>$ and $\sigma^2$ and thus can be utilized in applications based on orbitronics.

*Electronic mail: dipten@cgcri.res.in

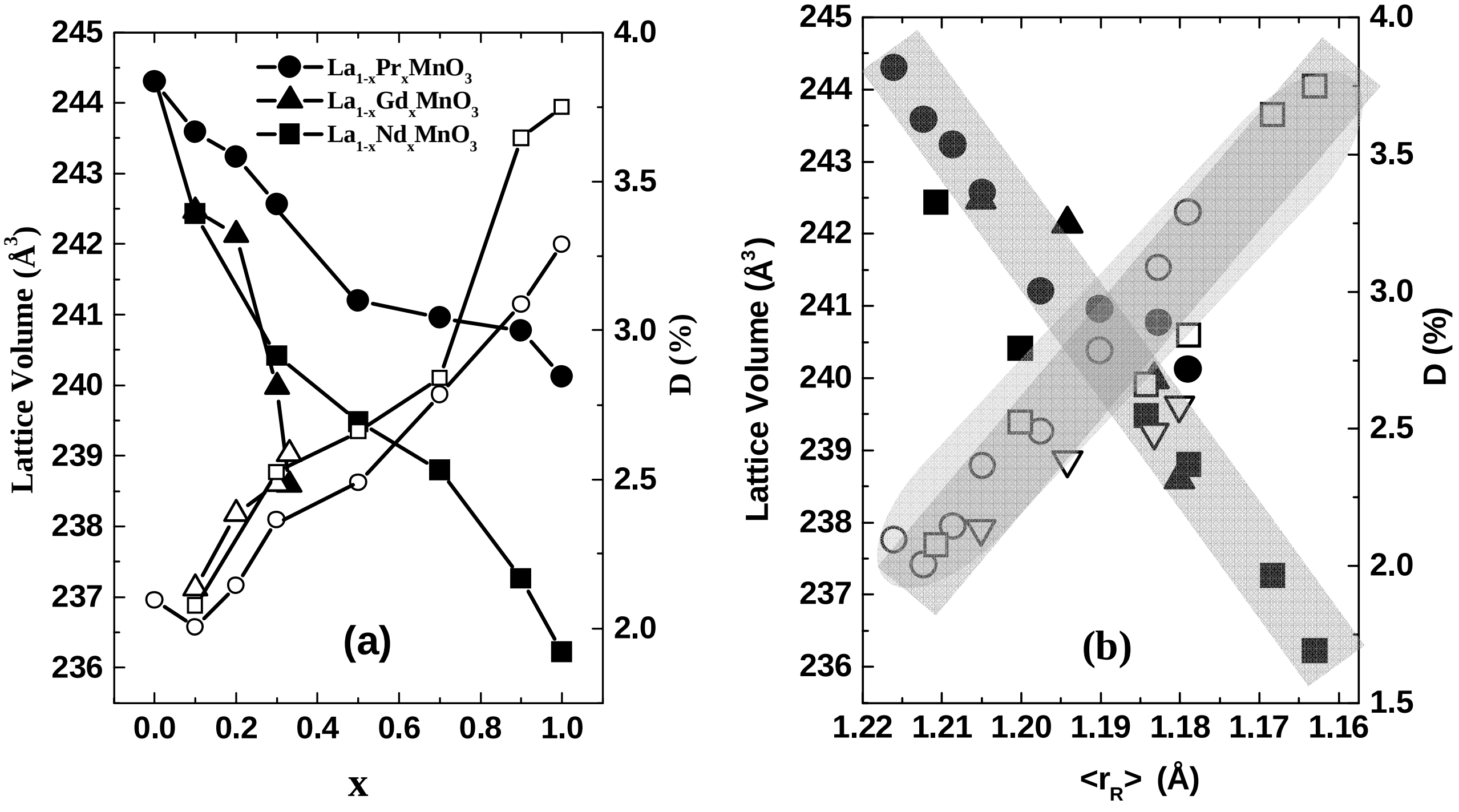

Fig. 1. Room temperature structural data for $La_{1-x}Pr_xMnO_3$, $La_{1-x}Nd_xMnO_3$, and $La_{1-x}Gd_xMnO_3$: (a) the lattice volume (solid symbols) and orthorhombic distortion (open symbols) as a function of x and (b) the lattice volume and orthorhombic distortion as a function of $<r_R>$.

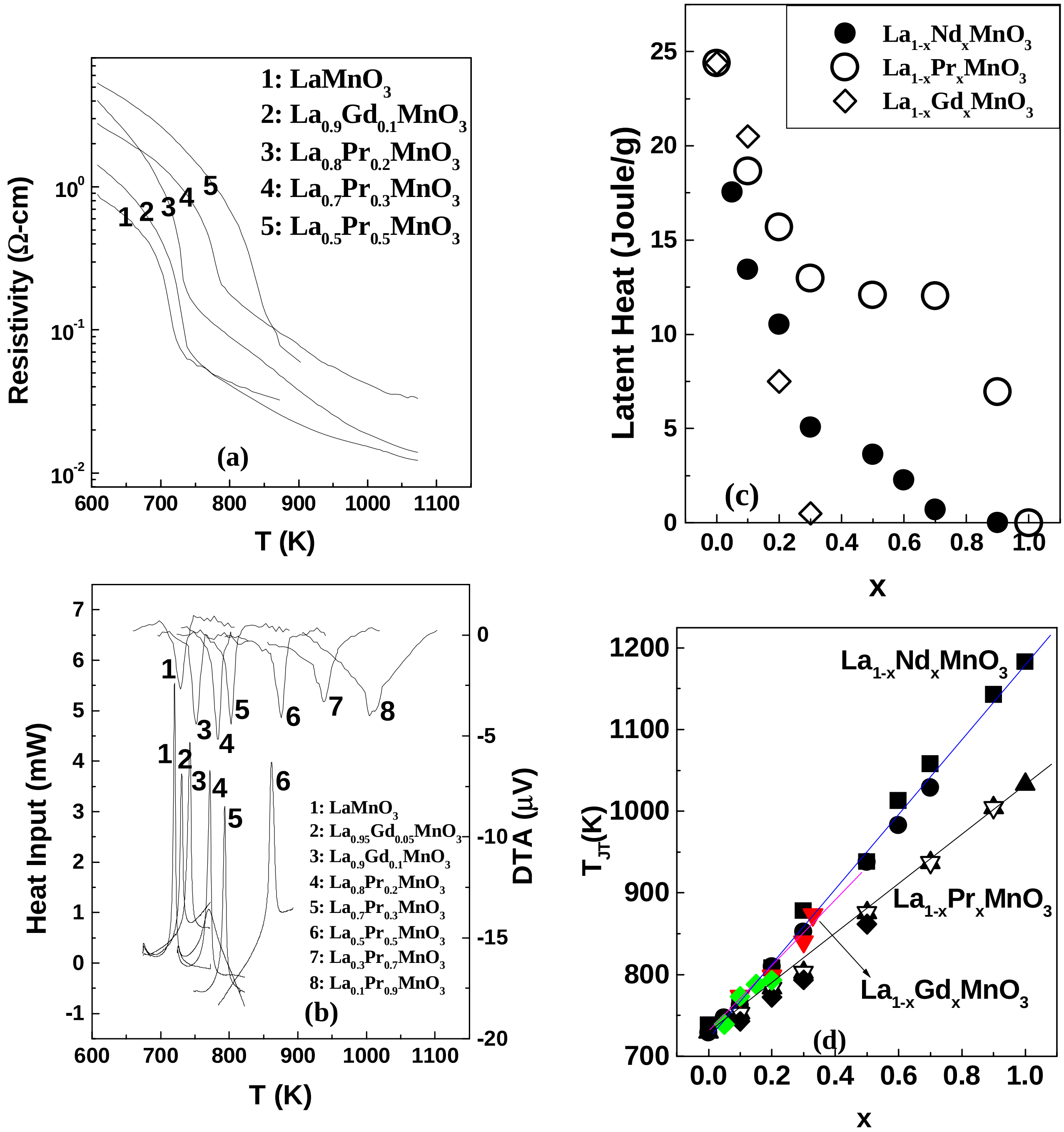

Fig. 2.(color online). The typical dc resistivity versus temperature plots (a) and the DSC and DTA thermograms (b) for few representative samples; (c) the latent heat versus x patterns for all the three series $La_{1-x}R_xMnO_3$ (R = Pr, Nd, Gd) and (d) the corresponding $T_{JT}$ versus a plots.

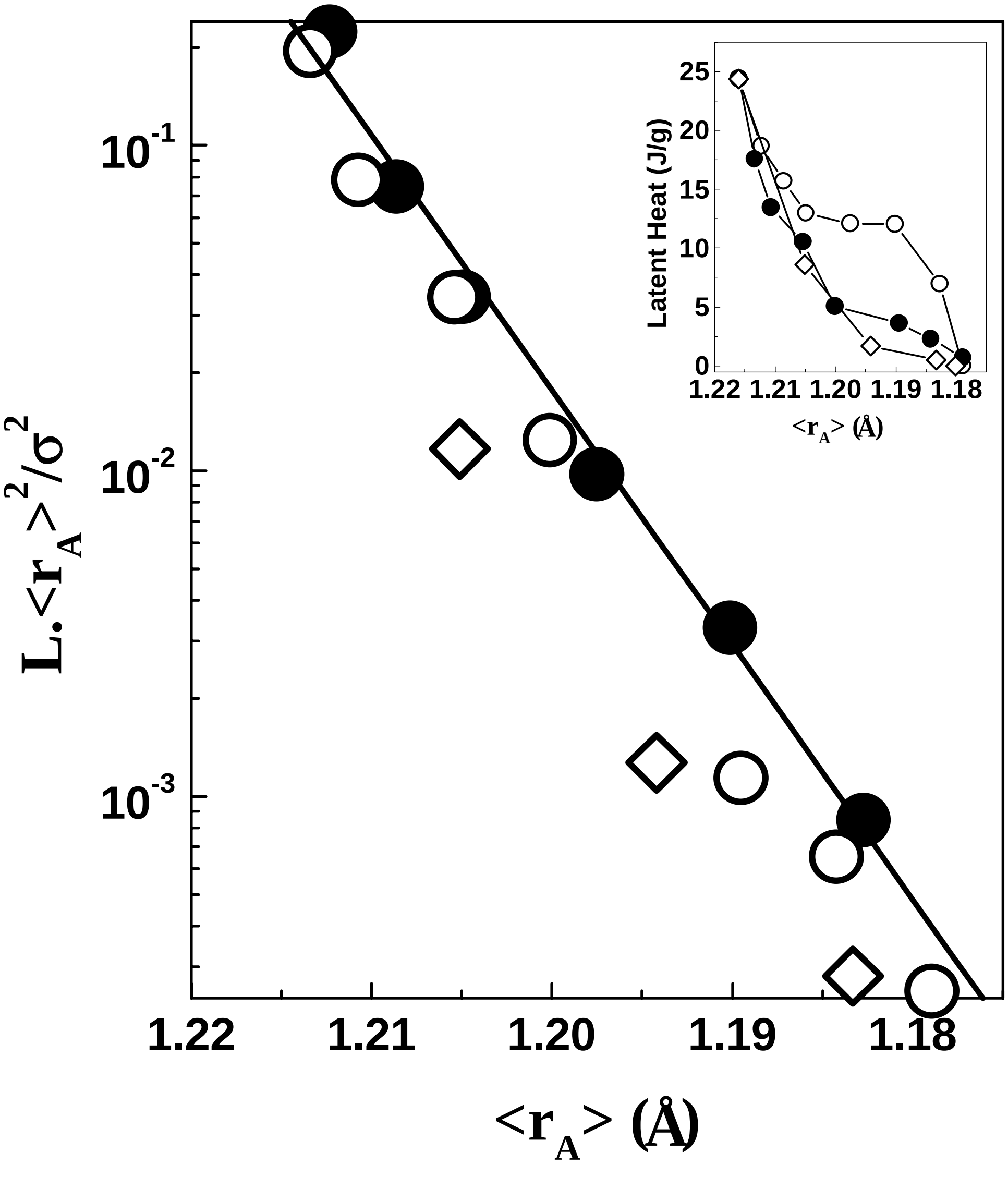

Fig. 3. The universal scaling of $L.\langle r_R\rangle^2/\sigma^2$ with $\langle r_R\rangle$: $L.\langle r_R\rangle^2/\sigma^2 \sim A.\exp[B.\langle r_R\rangle]$, $A = e^{-95.6}$ and $B = 78.2$; inset shows the latent heat versus $\langle r_R\rangle$ plot for all the three series of compounds.